\documentclass[preprint]{aastex}

\def\sun{\hbox{$\odot$}}
\def\etal{\it et al. \rm }
\begin{document}

\title{On the Structural Differences between Disk and Dwarf Galaxies}

\author{James M. Schombert}
\affil{Department of Physics, University of Oregon, Eugene, OR 97403;
js@abyss.uoregon.edu}

\begin{abstract}

Gas-rich dwarf and disk galaxies overlap in numerous physical quantities that make their
classification subjective.  We report the discovery of a separation between dwarfs and disks
into two unique sequences in the mass (luminosity) versus scale length plane.  This provides an
objective classification scheme for late-type galaxies that only requires optical or near-IR
surface photometry of a galaxy.  Since the baryonic Tully-Fisher relation for these samples
produces a continuous relation between baryonic mass and rotational velocity, we conclude that
the difference between dwarfs and disks must be because of their distribution of stellar light
such that dwarfs are more diffuse than disk galaxies.  This structural separation may be due to
a primordial difference between low and high mass galaxies or produced by hierarchical mergers 
where disks are built up from dwarfs.  Structural differences between dwarf and disk
galaxies may also be driven by the underlying kinematics where the strong rotation in disks
produces an axial symmetric object that undergoes highly efficient star formation in contrast
to the lower rotation, more disordered motion of dwarfs that produces a diffuse, triaxial
object with a history of inefficient star formation.

\end{abstract}

\section{INTRODUCTION}

Gas-rich galaxies dominate the far end of the Hubble scheme and are primarily composed of two
types, dwarf irregulars and spirals disks.  While differing in general morphological
appearance, the late-type dwarfs and disks also separate in terms of mean luminosity and size.
However, these galaxies are not uniquely determined by their mass or size as there exist many
examples of massive irregulars (Hunter \& Gallagher 1986) as well as dwarf spirals (Patterson
\& Thuan 1996, Schombert \etal 1995).  To first order, their kinematics are similar, both being
dominated by rotation (van Zee, Salzer \& Skillman 2001) and their star formation histories are
active both today and in the recent past (McGaugh \& de Blok 1997).

A connection between dwarfs and disks is also inferred by the baryonic Tully-Fisher relation
(McGaugh \etal 2000).  The constant slope and low scatter of dwarfs and disks in the
mass-rotational velocity plane implies that the corrections from luminosity to mass (IMF,
$M/L$, gas content) are similar from dwarfs to disks and that the total baryonic mass is
directly proportional to the total mass.  So, while the dark matter fraction may be higher in
dwarfs compared to spirals, this difference scales evenly across the Hubble sequence.  This
has implications for galaxy formation scenarios as hierarchical models, which would not
necessary predict two families of galaxies, would be supported by the discovery of distinct
differences in the properties of low and high mass galaxies (Kauffmann \etal 2003).

Given their similarities, the operative question becomes is there any difference between disks
and dwarfs or are they simply a continuous sequence where some underlying physical parameter
(such as angular momentum or gas fraction) determines the size and appearance of a gas-rich
galaxy.  The goal of this paper is to compare the properties of dwarf and disk galaxies by
fitting their surface brightness profiles in order to extract total luminosities, central
surface brightnesses and scale lengths in the search of structural properties which may
separate disk galaxies from dwarf irregulars.  We will also examine the H\,I properties of the
two samples (such as line profile shape and gas fraction) for details concerning kinematics and
star formation history that may suggest different paths of evolution between dwarfs and disks.

\section{ANALYSIS}

The data for this paper derives from three studies on the optical and H\,I properties of dwarf
and disk galaxies.  For the disk samples, we have used the data on ordinary spirals extracted
from two surveys, Courteau (1996) and de Jong (1996).  The Courteau survey was primarily
focused on Sc galaxies, as probes to the Tully-Fisher relation, and presented CCD and H\,I
observations of 189 galaxies.  The de Jong sample contains a range of spiral types selected
from the UGC with diameters greater than two arcmins and undisturbed in their morphology, a
total of 62 galaxies.  The dwarf sample of 107 galaxies is extracted from the LSB catalogs of
Schombert \etal (1997).  Both the optical and H\,I properties of these three samples were
discussed in detail in Schombert, McGaugh \& Eder (2001).  That paper focused on the gas mass
fractions of disk and dwarf galaxies in order to understand their star formation histories and
color evolution.  This paper will focus solely on the integrated optical and H\,I properties of
the three samples as well as their structural properties as given by surface photometry.  All
distance related quantities in this paper use values of $H_o = 75$ km sec$^{-1}$ Mpc$^{-1}$, a
Virgo central velocity of 977 km sec$^{-1}$ and a Virgo infall of 300 km sec$^{-1}$.  The data
used for this study is available at the LSB dwarf web site (abyss.uoregon.edu/$\sim$js).

The structural properties for both the disk and dwarf samples are determined by exponential
fits to their surface brightness profiles.  An exponential light profile is an almost universal
observational feature of disk galaxies (de Jong 1996) parameterized by two quantities, the
central surface brightness ($\mu_o$) and the disk scale length ($\alpha$).  Likewise, the
galaxies in the LSB dwarf sample are also well fit by an exponential light profile (Pildis,
Schombert \& Eder 1997), although there is no assumption that the shape of the underlying mass
distribution is identical to spirals (see section \S 2.1).  From these fits, the total
luminosity of the galaxy (integrated light based on elliptical apertures from the surface
photometry) is also extracted.  We use the $I$ band (9000\AA) observations for the surface
brightness profiles since the far-red provides a better measure of the total stellar mass
compared to near-blue colors.

\begin{figure}
\centering
\includegraphics[width=12cm, angle=-90]{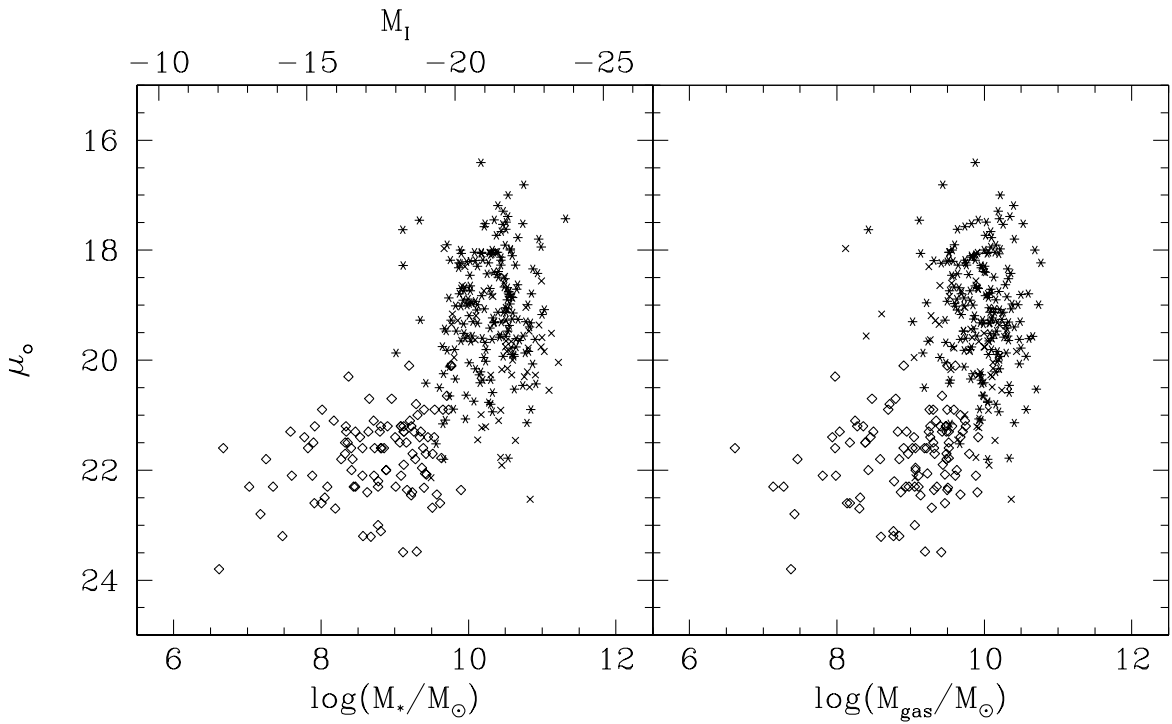}
\caption{
Stellar and gas mass versus central optical surface brightness ($\mu_o$).  The open
symbols are from the LSB dwarf catalog, crosses are disks from de Jong (1996) and asterisks are
Sc galaxies from Courteau (1996).  Luminosity is converted to stellar mass using an $M/L$ of
1.7 (see text).  Gas mass is determined from the integrated H\,I flux corrected for molecular
gas. The selection criteria for the LSB dwarf sample was low in mean surface brightness and
irregular in morphology.  This will naturally produce a sample of low central surface
brightness, low mass objects which occupy the lower lefthand portion of the diagram.  Disk
galaxies, having higher rates of past star formation, occupy the upper portion of the diagram.
Stability requirements keep their masses above those of dwarfs.}
\end{figure}

In studying structure of a galaxy, as given by its surface brightness profile, one is less
interested in the luminosity of a galaxy than its mass.  Gas masses can be determined directly
from 21-cm fluxes (with a small correction for the contribution from molecular gas).  However,
the conversion of luminosity to mass requires knowledge of the mean $M/L$ of the underlying
stellar population.  As discussed in Schombert, McGaugh \& Eder (2001), we follow the
prescription of McGaugh \etal (2000) and adopt a value of $M/L$=1.7 to convert luminosities
into solar masses.  The structural and mass properties of the three samples are summarized in
Figures 1 and 2, plots of galaxy mass versus central surface brightness ($\mu_o$) and scale
length ($\alpha$).  The galaxy mass is divided into two parts, the mass of stars ($M_*$) as
given by the $I$ band luminosity and the mass in gas ($M_{gas}$) as deduced from the H\,I
observations.  The sum of these two represents the total baryonic mass of a galaxy (McGaugh
\etal 2000).

Figure 1 displays the effect of a morphological selection combined with low surface brightness
criteria to the LSB dwarfs.  Both of the disk galaxy samples display a larger range of central
surface brightnesses than the dwarf sample reflecting their typically higher past rates of star
formation.  The disk samples also have higher mean masses as required by disk stability
requirements (Fall \& Efstathiou 1980).  While there is a trend of lower central surface
brightness with decreasing mass, this is, in part, due to how the samples were selected.  The
LSB dwarf sample was intentionally selected to be of lower surface brightness than objects in
the UGC catalog.  With the additional constraint of irregular morphology (Schombert, Pildis \&
Eder 1997), this criteria produces a sample of galaxies with low baryonic mass and low $\mu_o$.
The upper left corner of Figure 1 would be populated by blue compact dwarfs (BCD's, see
Patterson \& Thuan 1996) and dwarf ellipticals (dE's), but these types of galaxies are not
found in any of the three samples used herein.  There is no strict linear correlation in Figure
1 and galaxy formation models predict galaxies will occupy all regions of this diagram
(Dalcanton, Spergel \& Summers 1997) even if found to be rare in our galaxy catalogs.  The
diagram is more compact in the right hand panel for gas mass simply because of the gas-rich
nature of dwarf irregulars (i.e. more gas than stars).

\begin{figure}
\centering
\includegraphics[width=12cm, angle=-90]{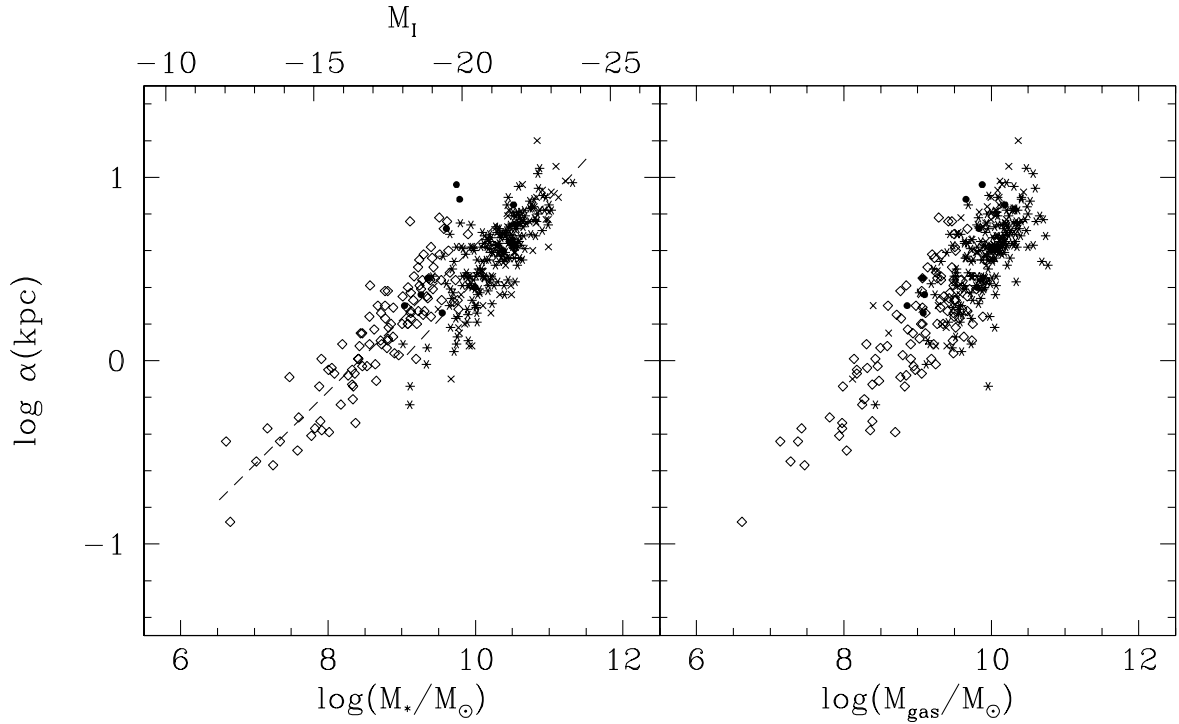}
\caption{
Stellar and gas mass versus optical scale length ($\alpha$) in kpc.  The open
symbols are from the LSB dwarf catalog, crosses are disks from de Jong (1996) and asterisks are
Sc galaxies from Courteau (1996).  The separation of dwarfs and disks into two sequences is
evident in the left panel.  Sm class galaxies from de Jong are shown as solid symbols and are
typically found on the dwarf sequence.  Biweight fits to each sample is shown as dashed lines.}
\end{figure}

Figure 2 displays the relationship between galaxy mass and scale length.  All three sets of
galaxies span five orders of magnitude in stellar mass and a factor of 50 in scale length.  The
obvious correlation is evident, i.e. larger galaxies are more massive.  However, it is also
clear from left-hand panel of Figure 2 that the data (in terms of stellar mass versus scale
length) form two separate and distinct groups.  One group is delineated primarily by the LSB
dwarfs and lies above (larger $\alpha$) and to the left (lower stellar mass) of the second
group outlined by the disk samples.  While there are a few disk galaxies located in the LSB
dwarf group, a majority of those galaxies are Sm class (see below).

Each group in Figure 2 is best described by a linear sequence.  The two sequences, which we
will refer to as the dwarf and disk sequences, are not a single track with a changing slope and
there is no evidence of any deviation from a straight line for each sample (see fits below).
There appears to be a break at $M_* = 10^{10} M_{\sun}$, but closer inspection shows that there
are several galaxies on the dwarf track of higher mass, and several spirals on the disk track
lower in mass but still following their respective sequences.  A biweight fit to the dwarf
sample results in the following relation:

$$ {\rm log}\, \alpha = 0.40 \pm 0.02\, {\rm log}\, M_*/M_{\sun} - 3.37 \pm 0.15 [dwarfs] $$

A biweight fit solely to the disk samples results in:

$$ {\rm log}\, \alpha = 0.44 \pm 0.02\, {\rm log}\, M_*/M_{\sun} - 3.96 \pm 0.19 [disks] $$

\noindent where both fits are shown in Figure 2 as dashed lines.  Inspection of the dwarf
galaxy data finds all the dwarfs to lie above the disk relation.  This cannot be due to a
change in slope for the dwarf sample as the linear fit to the dwarfs is statistically the same
slope as the disk sample.  The scatter around the linear fits is similar for the dwarf and disk
samples, even though the disk sample represents a brighter and photometrically more accurate
dataset.  This is probably a statement concerning the meaning of the parameters extracted from
surface photometry fits.  While both dwarfs and disks are well fit by an exponential for some
portion of their surface brightness profiles, the existence of bulges and irregular
luminosities distributions will reflect as variations in the total magnitude as compared to a
structural parameter such as scale length ($\alpha$) in Figure 2.  A disk or integrated
magnitude based on the surface brightness profiles produces less scatter, but than fails to
reflect the total stellar mass of a galaxy (see Graham \& Driver 2005).  There is no indication
that the scatter in Figure 2 disguises a non-linear correlation between $\alpha$ and stellar
mass, and the mean scatter is less than the separation of the two dwarf and disk sequences.
When examining the distributions of the disk and dwarf data around the linear fits, and using a
$t$-test to compare the resulting histograms, we find that we must reject the hypothesis that
the two samples derive from a common population at the 99.99\% confidence level.

The raw interpretation for Figure 2 is that LSB dwarf galaxies either have large scale lengths
for their stellar mass, compared to disks, or they have less stellar mass compared to disks of
the same scale length (i.e. they are too big for their mass or under mass for their size).  The
slopes of the two sequences are statistically indistinguishable from a value of $0.4$.   We
note that this produces a relationship of $L \propto \alpha^2$ and, therefore, given that the total
luminosity of an exponential profile is $L = 2\pi\alpha^2\Sigma_o$, this implies that the
optical luminosity of a galaxy will be independent of their surface brightness (see
Hoffman \etal 1996).  This is certainly true of LSB galaxies, where a range of luminosities,
masses and sizes is found from the LSB dwarfs to the Malin giants.

While the relationship between stellar mass and scale length ($\alpha$) differs between disks
and dwarfs, the same is not true with respect to gas mass.  The right panel in Figure 2
displays the relationship for gas mass, and the two samples overlap into one continuous
population.  This is due, primarily, to the differences in gas content between spiral disks and
LSB dwarfs.  LSB dwarfs have more of their baryonic mass in the form of gas compared to disks.
Thus, the disk sample decreases in the ratio of gas mass to stellar mass compared to dwarfs and
the two samples converge.

An interesting morphological point, there are several disks from the de Jong study on the dwarf
track, but defined as disks based on their morphological appearance (solid symbols in Figure
2).  All of these galaxies are Sm class, whereas the spirals on the disk track are classed as
Sd or earlier (see \S 2.1).  None of the Courteau Sc galaxies appear on the dwarf sequence.
Previously, Patterson \& Thuan (1996) have shown that dwarf spirals link, structurally, to LSB and
ordinary disks. The inclusion of these objects would extend the disk sequence to approximately
$M_T = -17$ or stellar masses of $5\times10^8 M_{\sun}$.

Why this separation of dwarfs and disks in the $M_*$-$\alpha$ diagram has gone unnoticed is not
clear.  The sequence is obvious in Patterson \& Thuan's data (see their Figure 10);
however, their interpretation is focused on the difference between gas-rich dwarfs (dI), dwarf
ellipticals (dE) and blue compact dwarfs (BCD), and the possible evolutionary connections
between those types.  In fact, the dwarf ellipticals appear to form a third sequence, fainter
for their size compared to spirals (or larger for their luminosity).  The two sequences are
also visible in Figure 2 of McGaugh \& de Blok (1997), and they even comment on the
discrepancy for the Sm/Im galaxies in the sample from the relationship in the $L_B-\alpha$
plane for Sa through Sd type galaxies.  Perhaps one reason that this separation is more obvious
in Figure 2 is that this study introduces another 5 magnitudes of dynamic range for the dwarf
sequence as compared to previous work and surveys strictly LSB galaxies.  HSB type dwarfs, such as
BCD's, appear to be related to dE's (Papaderos \etal 1996) and blur the relationship as they
lie on the disk sequence.

One explanation for the separation of dwarfs and disks in the $M_*$-$\alpha$ diagram would be
an error in the conversion from light to mass (i.e. the assumed $M/L$).  In order to bring the
two sequences into alignment, an error in $M/L$ between dwarfs and disks is required in the
direction that the dwarfs must have higher $M/L$'s by a factor of 8.  This is not plausible for
a number of reasons.  First, the luminosities selected to convert to mass are $I$ band
(far-red) measurements.  Stellar population models (Worthey 1994) demonstrate that $M_*/L$
varies with time and star formation rate as a function of wavelength, but is most stable in the
far-red because of its distance in wavelength from the region around the 4000\AA\ break.  Thus, $I$
band measurements provide a more accurate estimate of the stellar mass of a galaxy and obtain a
luminosity measure that vary little with recent star formation.  Second, this effect was
examined in detail in McGaugh \& de Blok (1997) who concluded that LSB galaxies cannot vary by
more than 20\% in $M/L$ as adjusted for color effects.  Lastly, a study by Bell \& de Jong
(2001) indicates that $M/L$ can vary by as much as a factor of two over the color range of LSB
dwarfs, but this is still well short of the factor of 8 required to align the dwarf and disk
sequences in Figure 2.

Another possible explanation to consider is a mismatch in the meaning of scale length from disk to
dwarfs.  The exponential profile is a good fit to rotating disks for kinematic reasons.  But
the justification for an exponential profile in dwarf irregular galaxies is solely based on the
empirical quality of the fit to the surface brightness profiles of dwarfs (Patterson \& Thuan
1996).  This is particularly true in the $I$ band where structural parameters (such as scale
length) are undistorted by recent star formation events or dust and provide a smoother measure
of the underlying stellar mass distribution.  However, one concern is that $\alpha$ is
measuring some quantity in 3D space that is not reflected into the sky distribution in the same
manner between dwarfs and disks.  This, then, concerns the true shape of dwarfs and disks.

It was concluded from imaging work that disk galaxies are, in general, axisymmetric oblate
bodies (Lambas, Maddox \& Loveday 1992), although lop-sided systems are not rare.  The
ellipticity of dwarfs was more recently studied by Binggeli \& Popescu (1995) for Virgo cluster
dwarfs and Sung \etal (1998) for a sample of field UGC selected dwarfs.  While the number of
gas-rich dwarfs is low in a rich cluster environment, Binggeli \& Popescu found that Im/dI type
dwarfs were slightly rounder than late-type spirals.  The Sung sample was selected by
morphological criteria, similar to the types of dwarfs used on our sample, and they conclude
that dwarfs and disks divide into structurally distinct populations.  Late-type disks are found
to be oblate spheroids with axial ratios of 1:1:0.2, whereas, dwarf irregulars are triaxial
with mean axial ratios of 1:0.7:0.5.  However, even in the extreme case of spherical dwarfs and
oblate disks, this would only account for an increase into 2D scale length of a factor of 1.2.
The correlations in Figure 2 are separated by a factor of 2.5, which cannot be explained by a
mere shift in the shape of dwarfs.

\section{DISCUSSION}

\subsection{Morphological Differences Between Disks and Dwarfs}

The two sequences in Figure 2 provide a new view on galaxy type outside the usual morphological
definitions.  While, in the last section, we have referred to the two sequences as dwarf and
disk, this is only because of the fact that a majority of the dwarf galaxies lie on one
sequence and a majority of the disk galaxies lie on the other, where we take the general
definition of a dwarf as a small, low mass system and a disk is a oblate, rotating galaxy.
There is a small, but significant, overlap in physical parameters such luminosity (mass), size
(scale length) and mass density (surface brightness) between the two sequences.  Therefore,
there is no boundary of mass, size or density to classify a galaxy as a dwarf without applying
an arbitrary division (Sandage \& Binggeli 1984).  In fact, the greatest importance for the two
sequences in the $M_*$-$\alpha$ plane is that it allows for a new technique to classify galaxies
as either dwarf or disk, based on two physical parameters (luminosity and scale length).

Since the distinction used herein between dwarf and disk in Figures 1 and 2 is based strictly
on morphological appearance of the galaxy, and not any characteristic as defined by mass or
kinematics, it is worth examining the morphological characteristics of the three samples.  The
Courteau and de Jong samples use the Hubble system to select a set of disk galaxies.  For
Courteau, this is a sample of Sc galaxies that are, to a high probability, a uniform set of
rotating oblate spheroids with small bulges and a clear spiral pattern.  The sample from de
Jong covers a wider range of Hubble classes, selected to be undisturbed and, therefore, also
forms a subset of the best examples of their morphological classes.  The late-type galaxies in
all three samples can be broken down into three morphological types; those with bulges and LSB
disks (Sc and Sd class), those with apparent axial symmetry but no bulges (Sm class) and those
with no symmetry (Im and dI class).  The de Jong and Courteau samples contain only those
galaxies with axial symmetry (Sa to Sm), so a disk galaxy in Figures 1 and 2 refers to this
morphological definition.

The dwarf sample was also selected on morphological grounds and focuses on Im and dI class
objects (although it also contains a number of Sm type systems, see Schombert, Pildis \& Eder
1997).  We found in a previous paper (Eder \& Schombert 2000) that an optimal technique to
produce a sample of low mass, gas-rich, small-sized galaxies was to search for LSB objects with
irregular morphology (the goal was to map the large scale structure of dwarfs).  However, that
technique does not preclude the inclusion of LSB disks into the sample. For example, there are
several examples of Sm type galaxies in the LSB dwarf catalog that are neither low in stellar
mass (luminosity) or small in scale length, but have sufficiently irregular morphologies to
appear similar to dwarf irregulars.  The rationale for the use of irregular morphology to
locate dwarf galaxies is a practical one.  Low mass density implies irregularity, at least with
respect to the pattern of star formation (Sneiden \& Gerola 1979), thus irregular structure is
nominally associated with dwarf galaxies (unless the star formation is completed, as in a dwarf
elliptical).  Other sources of irregular morphology, such as galaxy-galaxy interactions, are
normally associated with strong star formation activity and other tidal signatures.  Of course,
irregular morphology by itself is not an exact predictor of low mass for there are many
irregular galaxies that are massive (Hunter \& Gallagher 1986).  Likewise, axial symmetry does
not immediate guarantee that a galaxy is massive as there are many spirals that are dwarf-like
in luminosity and size (Schombert \etal 1995).

The three samples provide a range of Hubble classes, with overlap in the Sm class between the
LSB dwarfs and de Jong's sample.  All the galaxies classified as Im or dI lie on the dwarf
sequence and all the galaxies classified from Sa to Sd lie on the disk sequence regardless of
the sample from which they are derived.  The transition objects, the Sm class, all fall on the
dwarf sequence regardless if they are from the de Jong sample or the LSB dwarf sample.  This is
surprising since while Sm's are irregular, they are not chaotic and are certainly disk systems
in terms of being oblate spheroids (Sandage \& Binggeli 1984, Sandage, Binggeli \& Tammann
1985).  This indicates that something more fundamental than just disk-like shape and rotation
underlies the division found in Figure 2.

We also note that the two sequences appear as one continuous sequence when H\,I mass versus
scale length is considered (right panel in Figure 2).  This may be because of the fact that an
optical measure of size is being compared to the gas mass (H\,I flux).  Dwarf galaxies are
known to be more extended in H\,I compared to their optical image (van Zee, Haynes \&
Giovanelli 1995).  If the scale length ($\alpha$) was determined from H\,I images for these
galaxies, it seems plausible that those values would be larger and the two sequences would
again be distinguishable.  Carignan \etal (1990) find that the typical H\,I extent of a dwarf
irregular is twice to three times that of its optical extent, while spirals are only 50\%
larger in H\,I.  If spirals have smaller H\,I distributions than dwarfs, then this would
account for the lack of separation for H\,I mass versus scale length in Figure 2.

\subsection{Kinematic Differences Between Disks and Dwarfs}

The underlying reason to the structural differences between dwarfs and disks may be due to
kinematic differences.  A first estimate of the kinematics in gas-rich galaxies derives from
the shape of their H\,I line profiles.  These profiles can be divided into two types, gaussian
and double-horned.  A galaxy with a flat or rising rotation curve, and a distribution of gas
that declines with radius in the form of a power law, will exhibit a H\,I profile that has two
peaks with a flat plateau region between them, the so-called double-horned profile (see
Giovanelli \& Haynes 1988).  Those galaxies lacking in rotation (or observed face-on) will
display a gaussian profile where its width reflects the internal velocity dispersion of the
system.

All late-type spirals (except for face-on systems) have double-horned profiles since they are
oblate rotators.  However, there is no a priori reason to believe that an irregular dwarf
sample would also be composed solely of double-horned profiles.  For example, their irregular
shape may be due to the distribution of gas and stars determined by pressure support from an
anisotropic velocity distribution.  To test for the frequency of rotation in LSB dwarfs, we
examined the H\,I profiles of the galaxies in the Eder \& Schombert sample.  Of the 106
galaxies in their sample, 67 (63\%) display clear double-horned H\,I profiles.  However, half
of the remaining 39 galaxies have profiles with line widths less than 50 km/sec.  A profile
width this narrow is impossible to distinguish between gaussian or double-horned at the
resolution used on Arecibo system.  Therefore, their classification is unknown.  The remaining
20 galaxies have resolved gaussian shaped profiles, but this is only 24\% of the total sample.
If dwarfs are nearly oblate, then this value is consistent with the remaining fraction of
profiles being gaussian because of their nearly face-on orientations to the Earth.  Thus, we
conclude that a majority of LSB dwarfs have kinematics were rotation is evident and it is
plausible that all LSB dwarfs have rotation kinematics.  This also agrees with high resolution
H\,I mapping of late-type galaxies (Swaters \etal 2002) who finds that, although their
rotation curves are flatter than spirals, there is no indication that dwarf galaxies are not
rotation supported.

This means that the separation between dwarfs and disks seen in Figure 2 is not due to a simple
shift in the internal kinematics from rotation to non-rotation, at least not with respect to
the kinematics of the gas. The stars may follow different kinematics (Hunter \etal 2002)
although the gas and stars have similar density distributions in LSB galaxies (McGaugh,
Schombert \& Bothun 1995).  However, there is one key difference between the rotation of dwarfs
and that of spiral disks, the rotation values for dwarfs are much lower than spirals.  In this
low rotation realm, turbulent motion can compete with rotation as the primary influence on a
galaxy's kinematics (Seiden \& Gerola 1979).  As turbulent motion increases, relative to
rotation, the structure of a galaxy begins to deviate from an oblate spheroid.  A triaxial
structure, as indicated by the Sung \etal (1998) study, then forms where the shape of a galaxy
begins to take on the a global structure influenced by a slightly anisotropic velocity
distribution.  It is possible that this change, from rotation dominance to a stronger
turbulent component, in the kinematics of dwarfs is reflecting into structure and is signaled
by the two sequences in the $M_*$-$\alpha$ diagram.  Unfortunately, this scenario provides no
explanation for why there are two distinct sequences rather than a smooth transition in
structure from an oblate shape to a triaxial one as one progresses to lower galaxy mass.

A higher ratio of turbulent motion to rotation would also explain the irregular morphology of
dwarfs.  In disk galaxies, the symmetric nature to rotation reflects in the axial symmetric
luminosity distribution, broken only by local regions of higher surface brightness star
formation.  The flat rotation curve produces the star formation patterns that define the Hubble
sequence.  However, H\,I mapping of dwarfs shows that their kinematics follow solid body
rotation (van Zee \etal 1997).  With solid body rotation dominating over most of the optical
region of a dwarf galaxy, random cloud motion can produce a chaotic luminosity distribution
rather than a spiral pattern.  In this sense, our results follow the scenario proposed by
Brosch, Heller \& Almoznino (1998) where the irregular morphology in dwarfs is not caused by
stochastic star formation, but rather, the irregular structure is because of an asymmetry in
the dark matter halo and reflecting an irregular kinematic distribution.

\subsection{Star Formation in Dwarfs and Disks}

In Schombert, McGaugh \& Eder (2001), we showed that the stellar mass to gas mass relation for
spiral disks is steeper than for LSB dwarfs.  A steeper correlation for disk galaxies implies
that they have been more efficient at converting gas into stars in the past.  This also agrees
with the observation that disk galaxies typically have a greater amount of their baryonic mass
in stars rather than gas (McGaugh \& de Blok 1997) and that dwarf galaxies have young spectral
signatures ($D_n(4000)$ and H$\beta$, Kauffmann \etal 2003) .

One possible interpretation to the change in the relationship between gas and stellar mass in
dwarfs and disks is to consider the effects that different kinematics, suggested in the last
section, has on star formation processes.  Star formation mechanisms are, basically, methods of
increasing local gas densities until the Jeans mass is reached.  At this point, fragmentation
occurs followed by collapse into protostars.  There are three primary methods to achieve this
state; 1) cloud collisions (Elmegreen 1990, Scoville \etal 1991), 2) density waves (Kennicutt
1998) and 3) induced by shocks from SN explosions or stellar winds (Elmegreen \& Parravano
1994).

Considering cloud collisions first, the formation of clouds in disk systems is primarily
confined to the plane of the galaxy.  Since rotation is dominant, clouds will tend to gently
merge (i.e. with velocities similar to their internal velocities) with other clouds on the
similar orbits, but with slightly different orbital eccentricities.  If in dwarfs, on the other
hand, clouds are dispersed in a more diffuse, triaxial environment under the influence of
turbulent motion, then the cross section for collisions is smaller in this environment and, when
they do collide, the collisions will tend to be head-on and serve to dissipate the clouds
rather than increase their densities.  Thus, star formation, averaged over the lifetime of
a galaxy, will be less efficient in dwarf galaxies as compared to disks.

Density waves in disk galaxies also serve to focus star formation events.  Clumping of
molecular clouds has been noted in several spiral galaxies (Vogel, Kulkarni \& Scoville 1988,
Rand 1993, Vogel \etal 1993), all of which lead to enhanced star formation.  In particular,
this style of star formation becomes an orderly function of the kinematics of the spiral, and
gas is converted into stars in an even, efficient manner.  In contrast, the velocity
distributions in dwarf galaxies may prevent the build-up of density wave patterns, and star
formation becomes a strictly local phenomenon (Hunter, Elmegreen \& Baker 1998) and local star
formation is a chaotic process, i.e. less efficient in converting the gas supply into stars.

For induced, stochastic star formation, the mechanism that triggers cloud collapse is a
super-bubble shock wave.  The shock waves arise from a combination of stellar winds originating
in young clusters and/or supernova explosions (Tenorio-Tagle \& Bodenheimer 1988, Miller \&
Scalo 1978).  In disk galaxies, pattern speed is important because if a cloud spends more than
$10^7$ yrs in arm region then the chances that a nearby explosion will trigger star formation
are high (Sleath \& Alexander 1996).  If there is a larger component of random motion in
dwarfs, than this lowers the probability that a cloud will be located near an explosion, and
therefore lowers the star formation efficiency compared to a disk galaxy with coherent motion.
H$\alpha$ and H\,I maps of dwarfs confirm this chaotic aspect to star formation in dwarfs
(Walter \& Brinks 1999).

In summary, regardless of the exact mechanism for star formation, it is clear that the rotation
of disk galaxies leads to a more coherent star formation pattern and more efficient conversion
of gas into stars.  In contrast, if LSB dwarf galaxies have a larger component of random motion
over ordered rotation, then they will lack any coherent structure other than that temporarily
provided by self-progating star formation.  This, in turn, leads to random star formation and
inefficient use of the gas supply.  The end result is a galaxy that has distributed its stars
over a wider area (i.e. more diffuse as shown in Figure 2) versus a spiral galaxy that has
focused a majority of their stars in the compact disk region.

\subsection{Dwarf Galaxy Formation}

Figure 2 indicates that, per unit of stellar mass, dwarf galaxies have larger scale lengths
than disk galaxies.  If both types of galaxies are oblate rotators, and a majority of their
stellar mass is in the shape of a flattened spheroid component, then the differences detected
in Figure 2 would indicate that dwarfs and disks follow a different scaling law that describes
how stellar matter is distributed in their gravitational wells.  Dwarf galaxies, as a class,
have more of their stellar material at larger radii compared to disk galaxies of the same mass,
which also produces lower central stellar densities (see Figure 1).  It is important to note
that this difference is not a continuous sequence in stellar mass as there are clearly two
distinct relationships for dwarfs and disks in Figure 2.  Thus, despite whatever morphological
criteria has led us to classify dwarfs and disk galaxies, there is also some underlying
difference in the way stellar matter traces the gravitational potential that separates the two
types of galaxies.  Dwarf galaxies are not simply scaled down versions of disk galaxies,
although the stellar light distribution in both types follows an exponential law.  These
structural differences may have been imposed at the time of galaxy formation.  For example, all
galaxies have substantial dark matter halos, but the ratio of dark to luminous matter is
particular high in LSB dwarfs (de Blok \& McGaugh 1997).  Thus, the dwarf and disk sequences
may represent two families of galaxies that follow different formation paths or different
coupling methods to the dark matter halos in which they reside.

Aside from morphology, gas-rich galaxies display a large degree of regularity in their global
properties (Roberts \& Haynes 1994).  The recent advent of large galaxy datasets (i.e. Sloan
Digital Sky Survey, SDSS) allows a broader comparison of galaxy characteristics, and in
quantitative ways.  Several recent studies have noted a difference in galaxy properties (such
as spectral and concentration indices) between high and low mass systems (Tremonti \etal 2004,
Kauffmann \etal 2003).   Low mass galaxies in the SDSS samples display younger stellar
populations, lower surface mass densities and lower mean concentrations (i.e. lower mean
surface brightnesses) than high mass galaxies.  The division between the two families occurs at
3$\times$10$^{10}$ $M_{\sun}$, which is effectively the upper limit of the dwarf sequence as
seen in Figure 2.  Their correlations between stellar mass and various spectral indices also
overlap in the region of 5$\times$10$^9$ $M_{\sun}$ to 5$\times$10$^{10}$ $M_{\sun}$, identical
to the overlap region between disk and dwarfs herein.

Structurally, the Kauffmann \etal SDSS study finds a continuous sequence from high to low mass
galaxies where the concentration index for low mass galaxies is lower than high mass galaxies
(more diffuse), but the distinction is not as clear as their spectral indices because of their
use of a mean surface brightness value rather a direct profile fit scale length (i.e.
$\alpha$).  In contrast, the results presented here demonstrate that our two families (disk and
dwarfs) overlap in size, but form two distinct sequences in the $M_*$-$\alpha$ plane.  We
also note that the distinction between the disk and dwarf sequences is less noticeable when
only the gas mass is considered (right panel of Figure 2), but this is a comparison between the
optical scale length ($\alpha$) and the H\,I gas mass.  Since the distribution of H\,I gas in
dwarfs is typically more extended than the optical light (van Zee, Haynes \& Giovanelli 1995),
then it is reasonable to assume that the two sequences would be clearer when using H\,I scale length
and, also, the total baryonic mass (stellar plus gas mass).

In addition to star formation history and structural differences, low mass galaxies also
display different mass-metallicity relations as compared to high mass systems (Garnett 2002).
Garnett's work finds a uniform change in $O/H$ abundance with mass for galaxies with rotational
velocities ($V_{rot}$) greater than 125 km/sec.  Lower mass galaxies have a steeper
mass-metallicity slope that is interpreted as an increased loss of metals from low mass
galaxies by galactic winds.  It will be a challenge to galaxy formation models to explain why
there are strictly two sequences.  One solution would be the hierarchical clustering and
merging scenario of galaxy formation (Kauffmann, White \& Guiderdoni 1993, Kauffmann \& Charlot
1998) where the disk sequence is built by accertion from dwarf galaxies.  However, this would
mean that the dwarfs are primordial, which is not true of a majority of their stellar populations.

\section{CONCLUSIONS}

The primary goal of this paper is to demonstrate that there is a clear distinction between
standard disk galaxies and irregular dwarf galaxies in the mass (luminosity) versus scale
length ($\alpha$) plane.  Whether a galaxy lies on the dwarf sequence is not solely determined
by either morphology, luminosity, size, density or kinematics.  Yet the combination of stellar
mass and scale length produces two sequences that are distinct and must reflect some underlying
physics that is unique to dwarf galaxies as compared to spirals.  In fact, the distinction
between the two sequences is clear enough to serve as a future definition of a dwarf galaxy (as
one on the upper sequence) and a disk galaxy (as one on the lower sequence) regardless of
appearance.

We summarize our conclusions as the following:

\begin{itemize}

\item[(1)] Even though there is no single optical characteristic that distinguishes dwarfs from
disk galaxies, both types are clearly separated in the $M_*$-$\alpha$ diagram (Figure 2).  Given
that the Tully-Fisher relation for these two samples of dwarfs and disks produces an extremely
tight relation between baryonic mass and rotational velocity (McGaugh \etal 2000), the
difference must be in how the scale length develops in dwarfs rather than any function of mass
(either stellar or gas).  Therefore, we conclude that the dwarf galaxies form a distinct
sequence that is more diffuse (larger scale length, $\alpha$, for its mass) than disk galaxies.
The disk galaxies define a similar linear sequence despite the vast range in Hubble type and
central surface brightness, $\mu_o$.

\item[(2)] While the morphological separation is not perfect in Figure 2, all Hubble types from
Sc to Sd lie on the disk sequence and all Hubble types Im and dI lie on the dwarf sequence.  A
majority of the class Sm galaxies lie on the dwarf sequence, despite their apparent oblate, disk
shape.

\item[(3)] Several recent studies of the SDSS dataset have noted a difference in galaxy
properties (such as spectral and concentration indices) between high and low mass systems
(Tremonti \etal 2004, Kauffmann \etal 2003). The division between the two families occurs at
3$\times$10$^{10}$ $M_{\sun}$, which is effectively the upper limit of the dwarf sequence as
seen in Figure 2.  The existence of two families of galaxies would support hierarchical
clustering and merging scenarios of galaxy formation (Kauffmann, White \& Guiderdoni 1993,
Kauffmann \& Charlot 1998) where the dwarf sequence evolves into the disk sequence by
accretion.

\item[(4)] The structural difference between dwarfs and disks may reflect small shape and
kinematic differences between the two populations.  If the stellar and gas distribution of
dwarfs are triaxial (Sung \etal 1998), then the only quantitative difference between the
kinematics of dwarfs and disks is the ratio of their large-scale turbulent motion to rotation,
where dwarfs have low rotation rates and stronger non-axial motion.  This would lead to a more
diffuse stellar distribution and irregular morphology.

\item[(5)] The importance of large-scale turbulent motion over rotation would also explain the
irregular star formation patterns in dwarf galaxies.  This would also disrupt the internal star
formation processes and explain the why dwarfs have been more inefficient at converting gas
mass into stellar mass.  However, this hypothesis is tentative as we would expect a smooth
transition from low mass dwarfs to higher mass disks.  Instead, the two sequences signal two
distinct paths of galaxy evolution, perhaps controlled by separate processes.

\end{itemize}

Existence of two families of gas-rich galaxies suggests that young galaxies follow at least two
paths during their primeval stage after collapse of the initial lump of dark and baryonic
material.  For example, high mass galaxies can dissipate and, by conservation of angular
momentum, become rotating disk galaxies.  What turbulent motion was contained in the original
protogalaxy is overwhelmed by dissipation into the ordered motion of rotation.  For low mass
lumps, rotational kinematics still dominate, but large-scale turbulent motion plays a minor,
but significant, role by slowing the collapse that, in turn, leads to a more diffuse stellar
distribution and a primeval dwarf galaxy with chaotic structure.  Thus, morphology follows from
a combination of internal kinematics and star formation history.  Likewise, as proposed by
hierarchical models, dwarf galaxies form first as irregular structures than merge into spiral
disks where dissipation introduces ordered motion and star formation.

The feedback from the kinematics to star formation becomes one of the relative importance of
ordered motion over disordered motions.  Ordered motion, as found in disks, leads to higher
star formation rates and more efficient conversion of gas into stars.  This efficient style of
star formation also feedbacks into the development of patterns and density waves, which guides
the morphological appearance of spirals.  Thus, the highly efficient disk systems develop into
early-type spirals with high star formation rates, high central surface brightnesses and a rapid
use of their gas supplies.  Less ordered systems become the late-type disk galaxies with more
fragmented morphological appearance dominated by bursts of star formation and early baryonic
blowout of metal enriched gas.  This scenario finds support from estimates of the gas depletion
timescales for spirals, which are less than a Hubble time but greater than that for dwarfs
(Schombert, McGaugh \& Eder 2001).

At the far end of this scheme lie the LSB dwarfs.  While they have rotation kinematics, this
rotation is primarily solid-body and, thus, there will be fewer cloud collisions that induces
cooling plus fewer patterns of density for gas to collect and less time spent in the regions of
shocks waves from recent star formation, which would induce new star formation.  Based on the
current low surface brightness nature of the stellar population, this inhibited star formation
history would stretch all the way back to the formation of the dwarf.  While this scenario
emphasizes the influence of internal factors on the evolution of a dwarf, environmental factors
would clearly have an enormous impact, especially in clusters; however, this cannot be tested
for in the current LSB dwarf sample as it is a pure field sample (Schombert \etal 1997).

\acknowledgements

We wish to thank the generous support of the Arecibo Observatory for the allocation of time to
search for HI emission from the candidate dwarf galaxies and Michigan- Dartmouth-M.I.T
Observatory in carrying out the photometry portion of this program This work is based on
photographic plates obtained at the Palomar Observatory 48-inch Oschin Telescope for the Second
Palomar Observatory Sky Survey, which was funded by the Eastman Kodak Company, the National
Geographic Society, the Samuel Oschin Foundation, the Alfred Sloan Foundation, the National
Science Foundation and the National Aeronautics and Space Administration.







\end{document}